# Weak links and phase slip centers in superconducting $MgB_2$ wires

**Maya Bar-Sadan, Gregory Leitus and Shimon Reich***

Department of Materials and Interfaces,
The Weizmann Institute of Science, Rehovot, Israel

**Abstract**

$MgB_2$ superconducting wires were produced by the Mg diffusion method. Scanning electron microscopy (SEM), optical microscopy, dispersive x-ray analysis (EDS) and XRD diffraction were used to study the physical structure and content of the wires. Magnetic properties ($Tc^m$, $Hc_1$, $Hc_2$, Jc by the Bean model) were obtained with a SQUID magnetometer, and transport properties ($Tc^r$, $Hc_2$, resistivity and residual resistivity ratio) were measured using a standard four-lead configuration. The V-I characteristics of the wires close to the critical temperature showed a staircase response, which was attributed to the presence of weak links, creating phase slip centers. The origin of those weak links is discussed in relation to their formation and structure.

* Corresponding author: Tel. +972-8-9342588
Email address: shimon.reich@weizmann.ac.il (S. Reich).

## INTRODUCTION

There is an agreement that grain boundaries in $MgB_2$ are transparent to current and there are no weak links associated with them. Transport measurements in bulk polycrystalline $MgB_2$ samples do not exhibit weak-link electromagnetic behavior at grain boundaries [1] or flux creep [2], phenomena which limit the performance of high-$T_c$ cuprates. However in the $MgB_2$ wires, prepared by diffusion of Mg into commercial boron wires, we observe weak links. These weak links cause discrete destruction of superconductivity in the wires, and this manifest itself as "steps" in the V-I characteristics. These steps can be attributed to the phase slip phenomenon.

In this paper we discuss the production of $MgB_2$ wires, their critical characteristics and the origin of the weak links vis-a vis the micro structure of the boron wires and the synthesis of the $MgB_2$.

## EXPERIMENTAL

Superconducting $MgB_2$ wires were produced by the procedure described by Canfield et al. [3]: The reaction involves 100 μm in diameter, commercially available, boron wires and the vapor of very pure magnesium metal at elevated temperatures. The reactants were sealed in a Ta tube in Ar atmosphere, sealed again in a quartz ampoule and placed in a preheated furnace (950°C). After 3 hours the quartz ampoule was removed and water quenched to room temperature. Taken that $MgB_2$ is the most Mg-rich binary Mg-B compound known, excess Mg was used to inhibit the growth of the higher boride phases, 4 times the proper stoichiometric ratio.

We characterized the composition of the wires by XRD spectroscopy: powder diffraction was obtained with a Rigaku RU-200 generator with a Rigaku D-MAX/B diffractometer and Kappa Nonius single crystal diffractometer with CCD detector, using the program for powder diffraction imitation. Magnetizations vs. field and vs. temperature were obtained with a Quantum Design $MPMS_2$ SQUID magnetometer. The resistivity and Volt-Ampere characteristics were measured in four points transport technique. The current was produced with a Keithley 224 programmable current source and the voltage was measured with a Keithley 182 sensitive digital voltmeter. Transport measurements were performed in the MPMS2 magnetometer utilizing its temperature and field controls.

**RESULTS and DISCUSSION**

The $MgB_2$ wires produced were brittle, deformed and had a rough surface which was probably the result of the Mg fast condensation during the quenching procedure. Estimation of the phase content from the surface of a bundle of wires, glued together by the Mg condensate, was carried out by an X-ray powder analysis. The estimation showed the presence of more than 60 vol.% of the $MgB_2$ phase together with Mg (>30 vol. %), and little portions of other phases, originating from the tungsten-boride core of the B fiber. The higher Mg-B phases ($MgB_4$ and $MgB_7$) were not detected. X-ray analysis of a single wire did not show any orientation preference of the $MgB_2$-grains and the wire did not seem to be highly crystalline. In fig.1 we show SEM secondary electrons images of the original boron and the $MgB_2$ wires, note the tungsten core in these images.
Side view (secondary and back scattered electrons) emphasizing the different composition of the roughness on the surface of the wire is shown in fig. 1c and fig.1d.
Side view and cross section EDS images showing the spatial distribution of Mg and B are presented in fig.1e,f,g,h. Note that Mg diffused almost up to the tungsten core. The condensation of Mg metal on the surface of the $MgB_2$ is observed in fig.1e.

The study of the magnetic and electro-transport parameters manifested superconducting properties, shown in fig.2 and summarized in table 1.

## Table 1: The critical temperature of the $MgB_2$ wires

| ***Magnetization*** | | | ***Transport*** | |
|---|---|---|---|---|
| *Susceptibility, emu/cc/Oe* | *On-set Tc, K* | *Transition width, K* | *On-set Tc, K* | *$\rho$(42K), $\mu$ohm-cm* |
| -0.052 | 39.0 | 3.5 | 41.0 | 12.9 |

Susceptibility values of wires from the same batch were in the range of -0.04 to -0.06 emu/cc/Oe with similar transition widths. The lower critical field, $Hc_1$, was estimated in the present study to be 140 Oe. The fitting parameter of temperature dependant resistivity, $\rho = \rho_0 + \rho^{\alpha}$ between Tc and 200K, $\alpha$ was ≈2.6.

The higher critical field, $Hc_2$, was defined first as field in which the magnetization signal merges with the background signal and vanishes. Transport isothermal curves were measured at different applied magnetic fields; $Hc_2$ was defined by the on-set criterion of 2% of the S.C. transition. These values are presented in fig.3. Both the critical temperature and the higher critical field were higher by the resistivity measurements, probably due to nature of measurement. The magnetization measurements include non-stoichiometric regions which are inferior in their superconducting character than the stoichiometric ones.

Using the Bean model [4] with $Jc=17\Delta M/r$, as it was proposed by Finnemore et al. [5] and used by Canfield et al. [3] for wires, the critical current densities can be derived from the hysteretic loops (fig. 4). Here Jc is in A/cm$^2$, ΔM in emu/cm$^3$, and the sample radius, r, in cm. The Bean critical current could be derived only for high temperatures, close to the transition, because a full loop must be obtained for that matter. The critical current density was also measured directly by a current-voltage curve, by a criterion of 1μV. The critical current density was calculated to be of the order of $10^4$ A/cm$^2$ at 35-36K (H=0) and in the order of $10^1$ A/cm$^2$ at 35-36K (H=5000-6000Oe), the same order of the critical current densities as those reported by Canfield et al.

The most interesting finding, which emerged from the above measurements, is the observation of a staircase response of the current-voltage characteristics, fig. 5 and fig.6. This response is suppressed by approaching the critical point, either by increasing temperature or by increasing the strength of the applied field. With an applied field of 5000 Oe, the temperature in which the "steps" vanish and the current-voltage characteristic becomes linear (or ohmic) is 37.5K, while it is around 40K without applied magnetic field. Another feature of these characteristics is the appearing of the "steps" at a specific current for each measurement, depending on the specific temperature and applied field in which the measurement was carried out.

The origin of the staircase response may be attributed to phase slip centers originating from weak links in the samples. This phenomenon was described in the following way [6]: when at certain location in the wire the critical current is somewhat smaller than elsewhere, an increase in current will bring this point to its normal state – causing normal flow of electrons through this weak point. This flow will induce an electric field. The latter, in turn, will accelerate the normal electrons to the critical velocity. Electron pairs at the weak spot will be broken, the amplitude of the order parameters $|\psi|$ will go to zero and the current will be carried by the normal component only. But at low enough temperatures the formation of electron pairs is favored. Therefore, the amplitude of the order parameters $|\psi|$ will become nonzero again and the current will be carried partly by S.C. electrons i.e. super current will appear. Then the entire process will repeat itself. On completion of each cycle, the phase difference of the S.C. electron wave functions to the left and to the right of the weak spot will vary by $2\pi$, so called phase slip center. The size of the center, $\delta$, is the region in which oscillations of $|\psi|$ take place, $\delta \approx 2\xi(T)$, where $\xi$ is the correlation length. Therefore experimentally the phase slip process manifests itself by a staircase response in the current – voltage characteristics, as discrete "steps" of resistivity are formed.

This response can be interpreted in view of the granular structure that is observed in the $MgB_2$ wires. During the synthesis the large volumetric expansion associated with the Mg diffusion causes cracks in the wires [7]. These cracks are formed at the interface between the corn like grains which are characteristic to the original boron wire [8]. The surface structure of the boron wire, see fig.7a, is associated with the relief of the substrate (tungsten core fiber) surface, determining the location of nucleation centers. From these centers boron grains grow radially in form of cones diverging upward and creating the corn like morphology on the surface of the wire. Corn like morphology is also observed in the $MgB_2$ wire, see fig.7b. Moreover, the wire probably contains different stoichiometric regions, demonstrated by the non-linearity of the M(H) curve and by lower $Hc_2^m$ (higher critical field from magnetization measurements) than $Hc_2^r$ (higher critical field from transport measurements). These regions may have different transport properties i.e. a difference of 8 orders of magnitude of the resistivity in the normal state, when the stoichiometric ratio changes to $Mg_{0.85}B_2$ [9]. As the transport measurements detect the current that flows through the most connected, stoichiometric route of grains, in a percolative model, when the connectivity is not sufficient, a weak link behavior is observed.

Acknowledgment: the Israeli Ministry of Infrastructure supported in part this research.

**FIGURE CAPTIONS**

***Fig. 1:*** SEM images of (a) the boron wire which serves as reactant and of the $MgB_2$ wire segment: SE image of a cross section (b) and of a side view (c). BS image of the side view emphasizing the different composition of the roughness on the surface and the wire (d) and EDS images showing the spatial distribution of Mg (e) and boron (f). EDS images showing the spatial distribution in a cross section of the wire of Mg (g) and boron (h).

***Fig. 2***: Zero Field Cooled (ZFC) and Field Cooled (FC) Magnetization (H = 50 Oe) of a single $MgB_2$ wire, located parallel to the field. ***Inset:*** *Resistivity vs. temperature at different applied fields of the same $MgB_2$ wire.*

***Fig. 3:*** Higher critical field, $Hc_2$ at different temperatures: Magnetization measurements (triangles), and by the on-set criterion from the isothermal transport measurements (circles).

***Fig. 4:*** Critical current density, Jc, at different applied fields and temperatures: By the Bean model (open symbols) and directly from the transport measurements with a criterion of 1 $\mu$V (closed symbols).

***Fig. 5:*** A typical V-I characteristics of a single wire of $MgB_2$, located perpendicular to the applied field. (a) At a constant temperature of 39K and different applied fields, from right to left: 0, 100, 200, 300, 400, 600, 800, 1000 and 2000 Oe. (b) Enlargement of the frame shown at (a).

***Fig. 6:*** V-I characteristics of a single wire of $MgB_2$, located perpendicular to the applied field. (a) Without applied magnetic field at different temperatures, from right to left: 39.0, 39.2, 39.3, 39.4, 39.5, 39.6, 39.7 and 39.8 K. (b) With an applied

magnetic field of 5000 Oe at different temperatures, from right to left: 35.00, 35.50, 36.00, 36.25, 36.50, 36.75, 37.00, and 37.50 K. (c) With an applied magnetic field of 6000 Oe at different temperatures, from right to left: 35.00, 35.50, 36.00, 36.25, 36.50, 36.75, and 37.00 K.

***Fig.** 7:* An optical microscopy image of the boron wire (a) and of a $MgB_2$ wire segment coated with an evaporated ~120 nm gold layer (b).

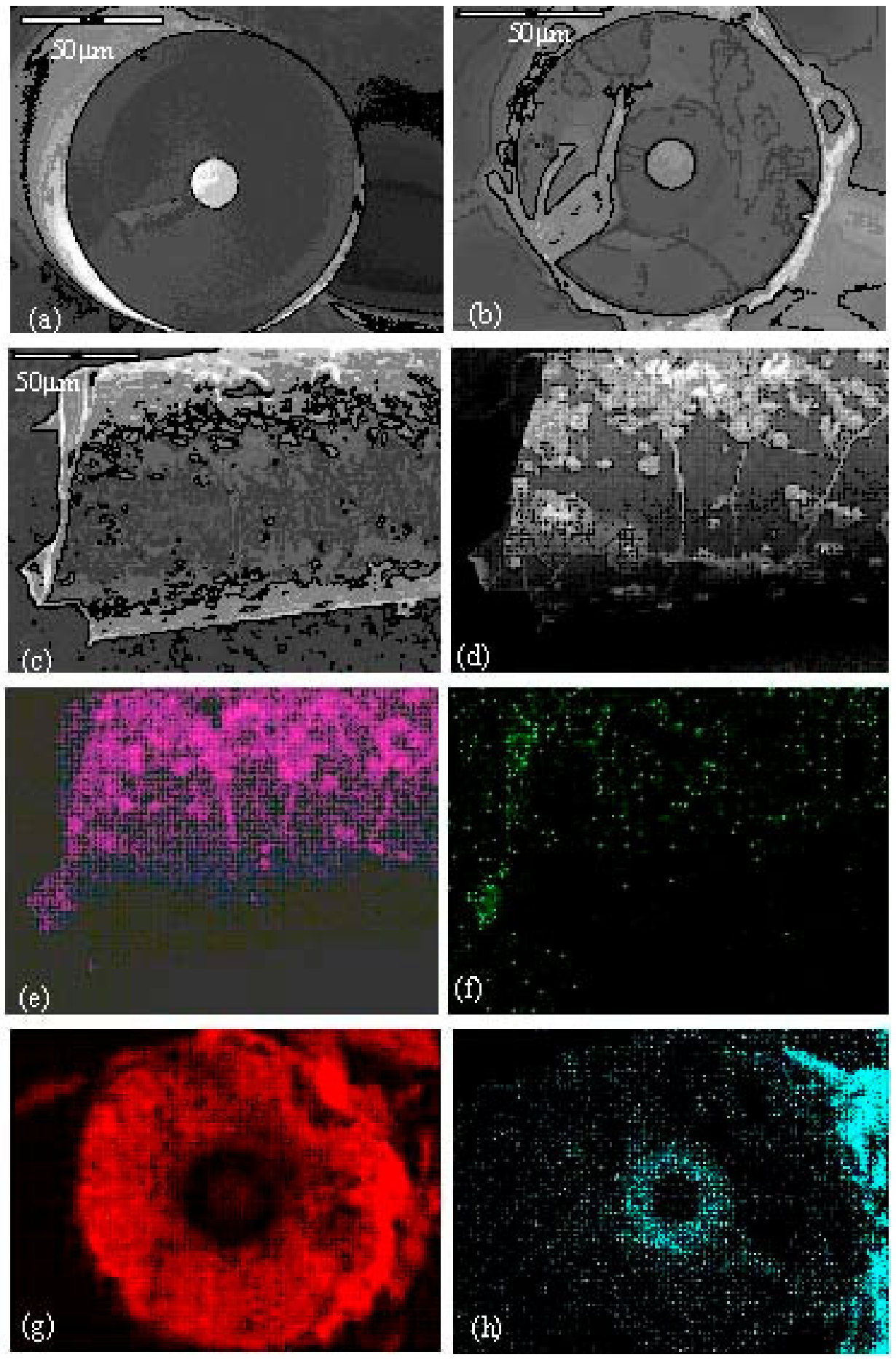

***Fig. 1:*** SEM images of (a) the boron wire which serves as reactant and of the $MgB_2$ wire segment: SE image of a cross section (b) and of a side view (c). BS image of the side view emphasizing the different composition of the roughness on the surface and the wire (d) and EDS images showing the spatial distribution of Mg (e) and boron (f). EDS images showing the spatial distribution in a cross section of the wire of Mg (g) and boron (h).

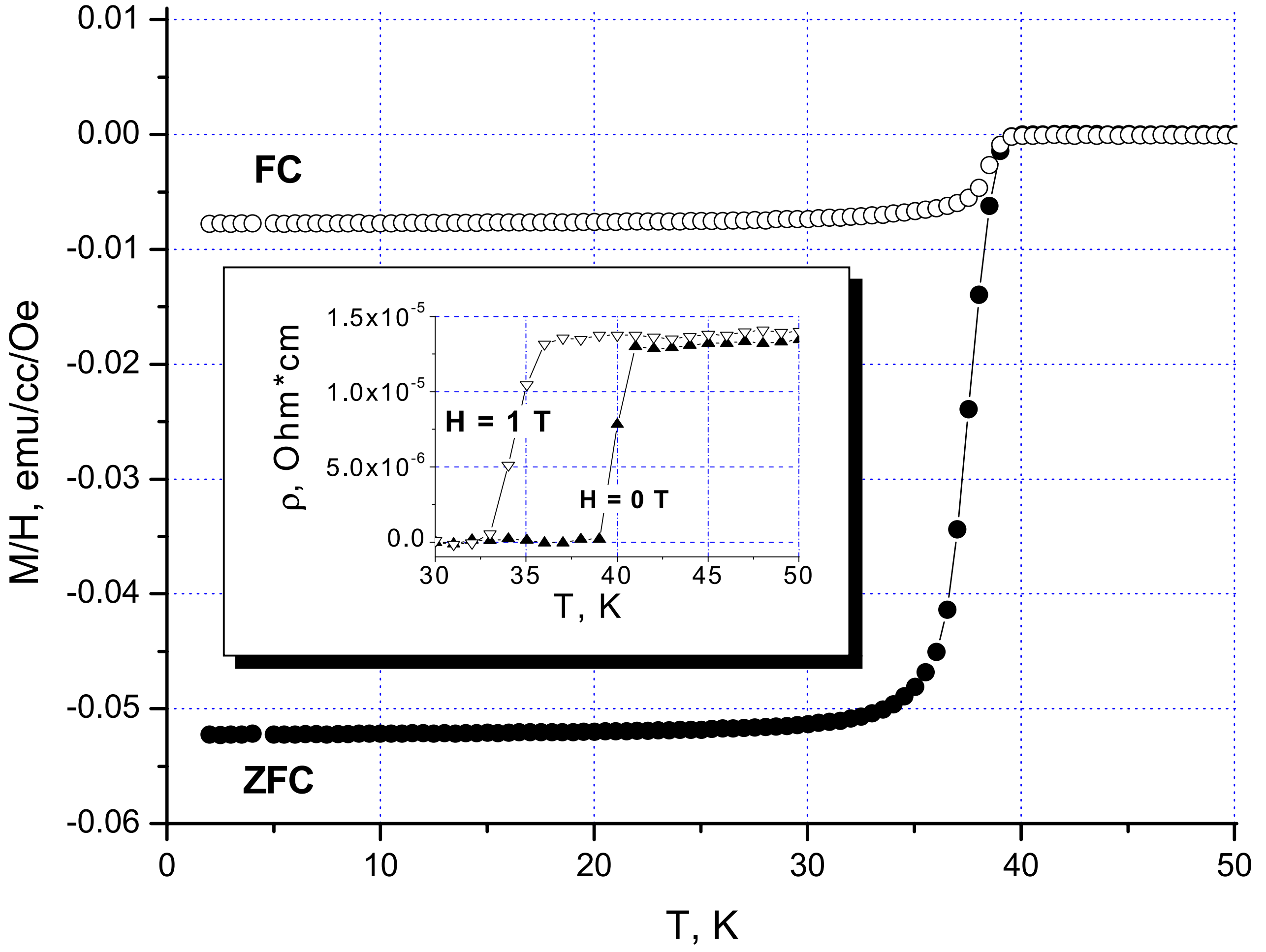

***Fig. 2***: Zero Field Cooled (ZFC) and Field Cooled (FC) Magnetization (H = 50 Oe) of a single $MgB_2$ wire, located parallel to the field. ***Inset:*** *Resistivity vs. temperature at different applied fields of the same* $MgB_2$ *wire.*

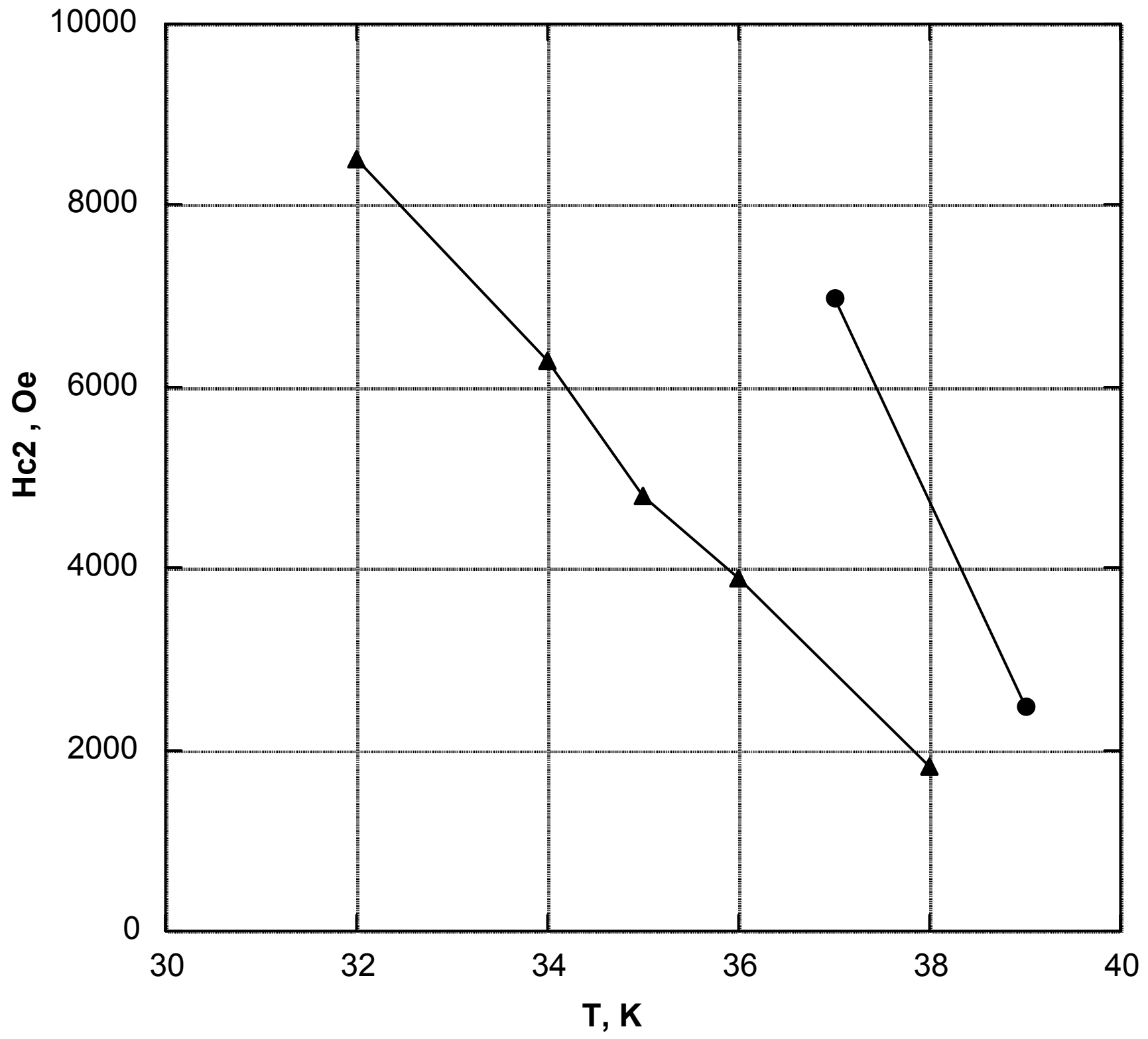

***Fig. 3:*** Higher critical field, $Hc_2$ at different temperatures: Magnetization measurements (triangles), and by the on-set criterion from the isothermal transport measurements (circles).

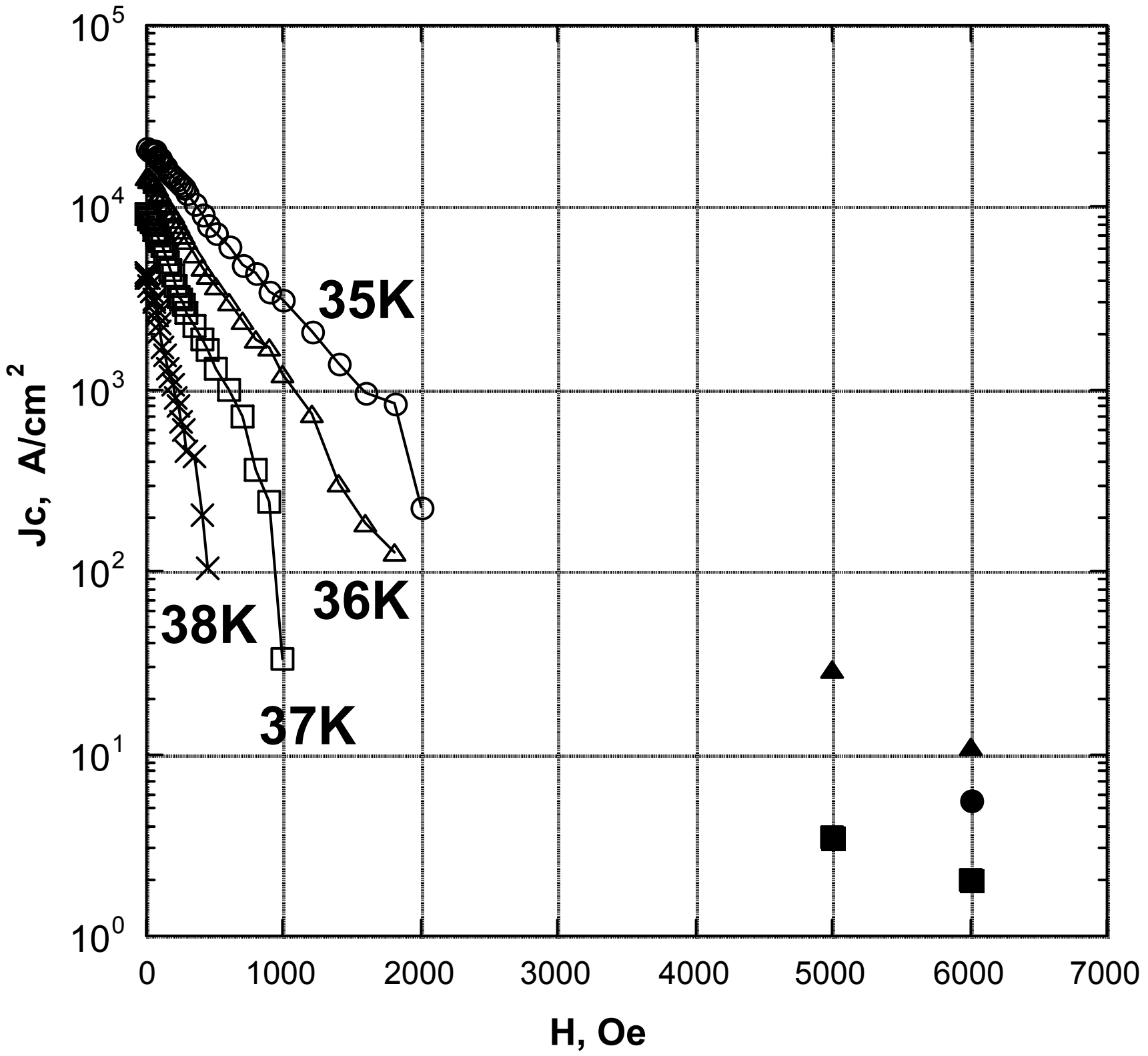

***Fig. 4:*** Critical current density, Jc, at different applied fields and temperatures: By the Bean model (open symbols) and directly from the transport measurements with a criterion of 1 μV (closed symbols).

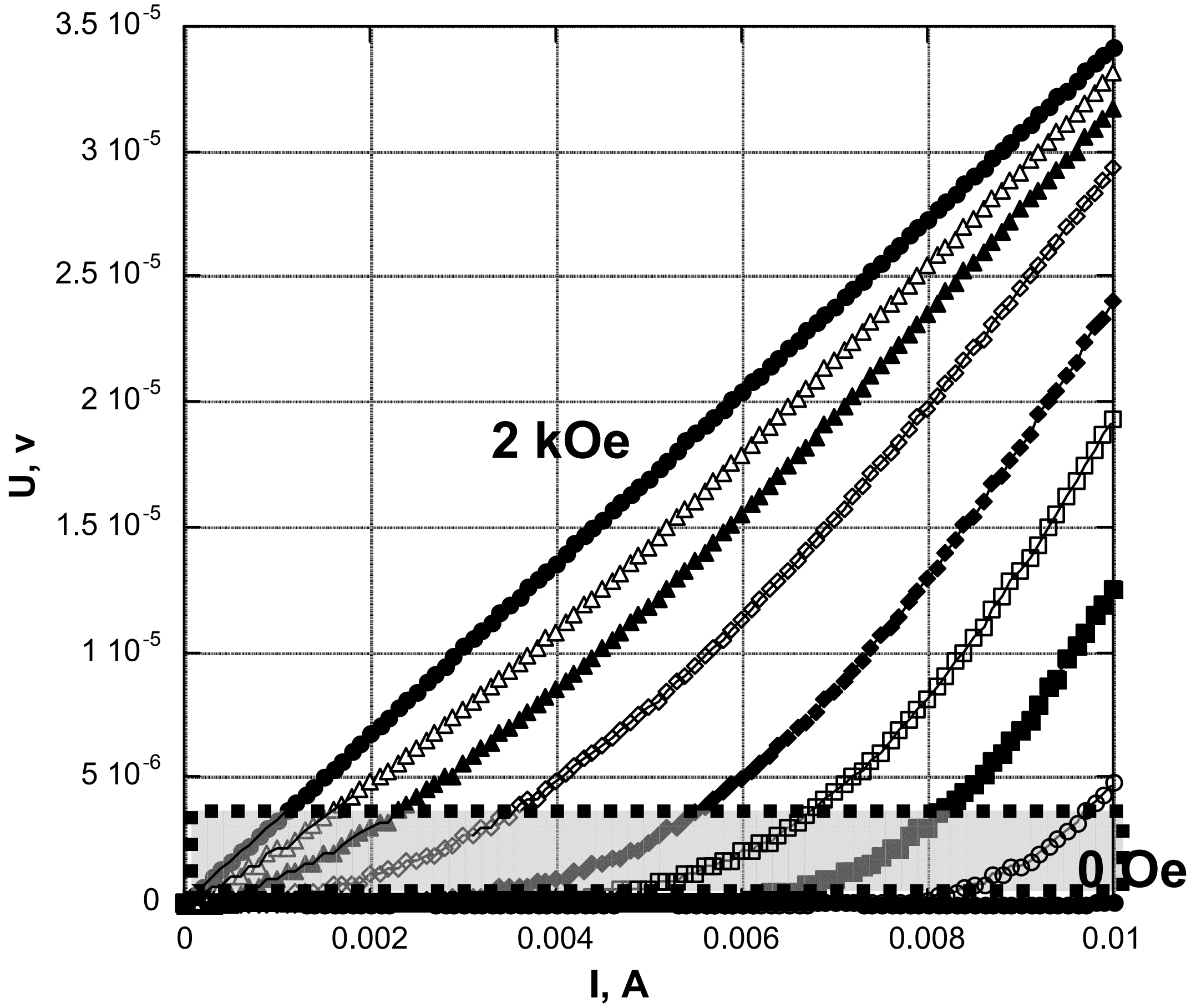

**(a)**

***Fig. 5:*** A typical V-I characteristics of a single wire of $MgB_2$, located perpendicular to the applied field. (a) At a constant temperature of 39K and different applied fields, from right to left: 0, 100, 200, 300, 400, 600, 800, 1000 and 2000 Oe. (b) Enlargement of the frame shown at (a).

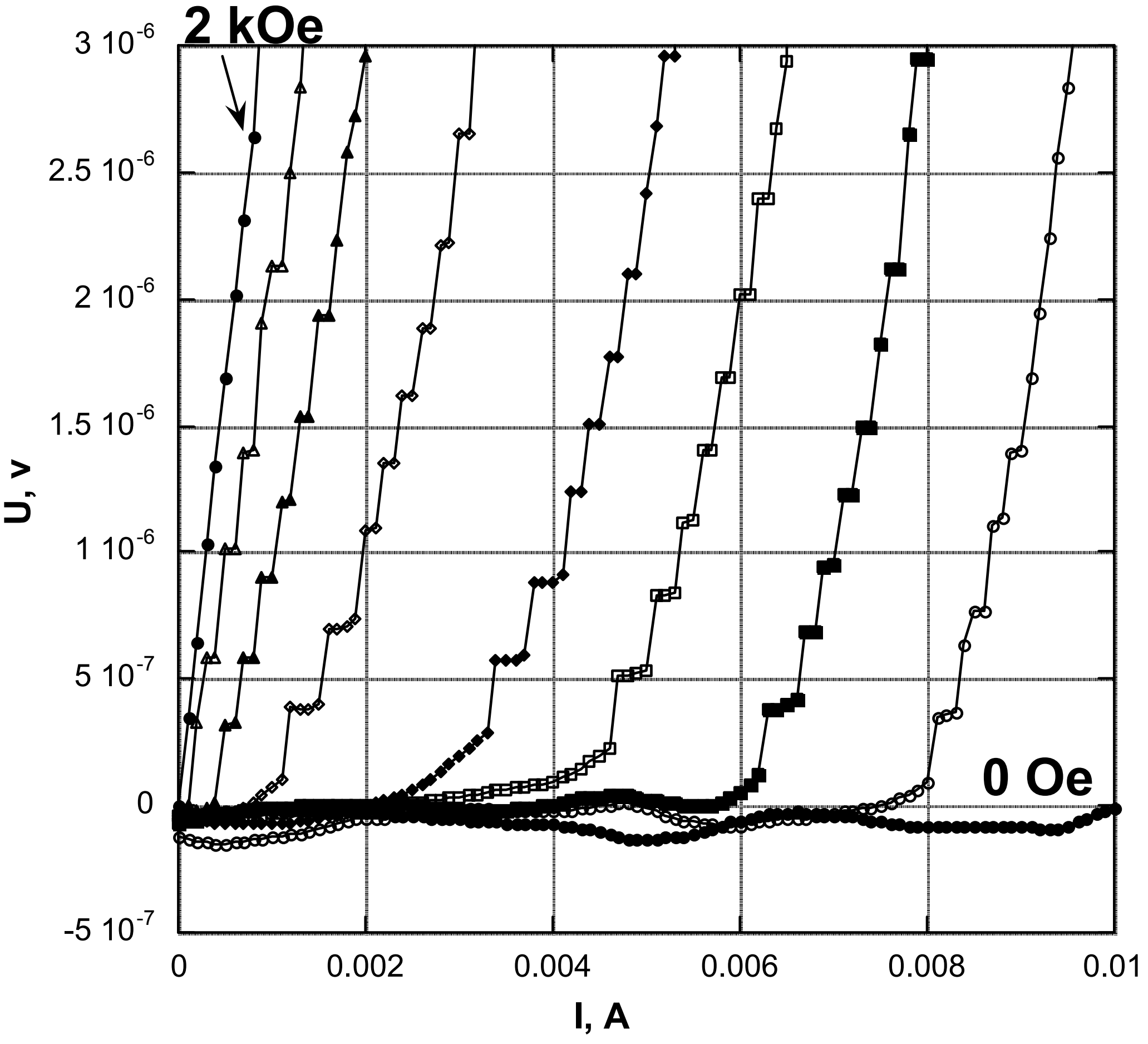

**(b)**

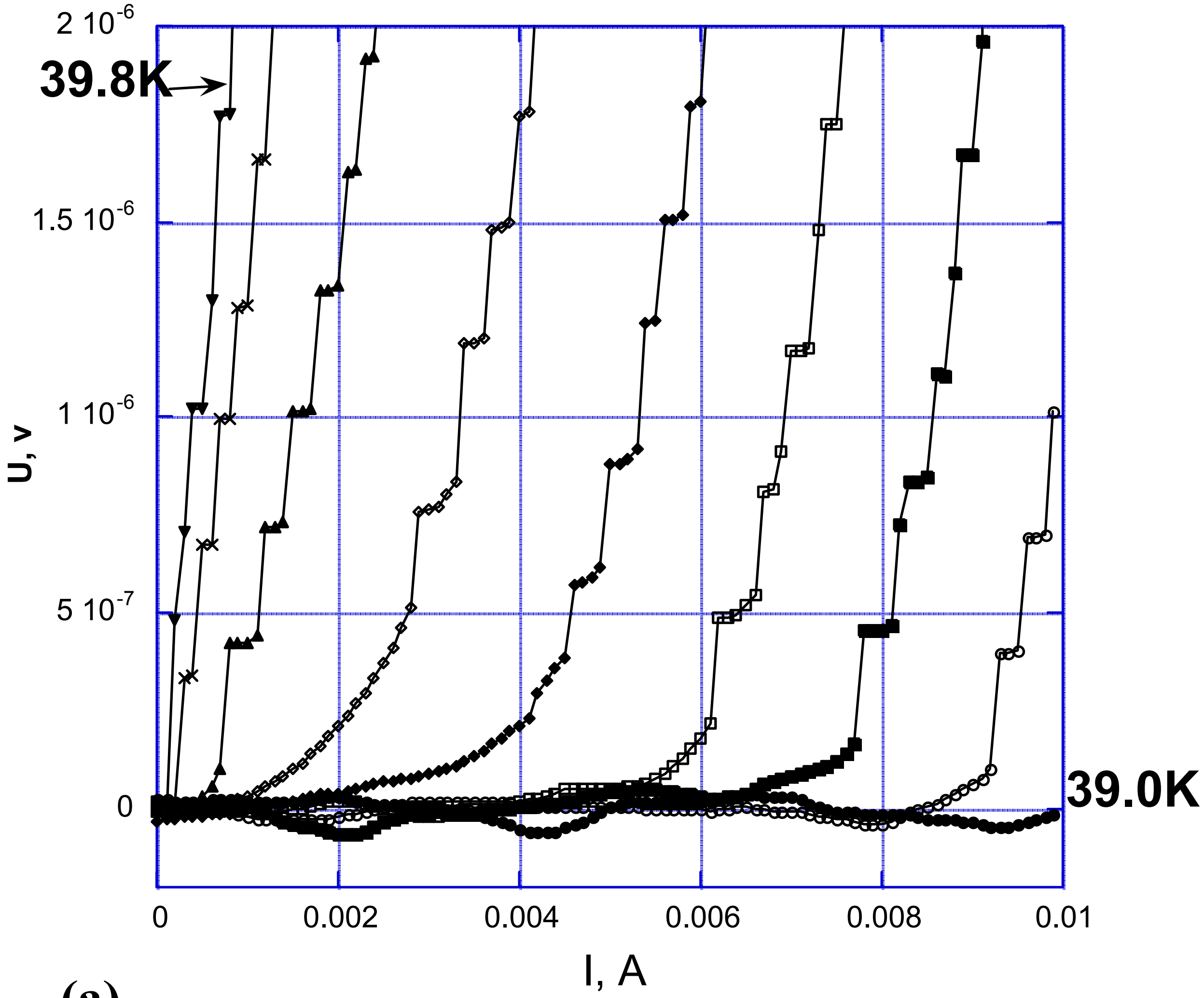

**(a)**

***Fig. 6:*** V-I characteristics of a single wire of $MgB_2$, located perpendicular to the applied field. (a) Without applied magnetic field at different temperatures, from right to left: 39.0, 39.2, 39.3, 39.4, 39.5, 39.6, 39.7 and 39.8 K. (b) With an applied magnetic field of 5000 Oe at different temperatures, from right to left: 35.00, 35.50, 36.00, 36.25, 36.50, 36.75, 37.00, and 37.50 K. (c) With an applied magnetic field of 6000 Oe at different temperatures, from right to left: 35.00, 35.50, 36.00, 36.25, 36.50, 36.75, and 37.00 K.

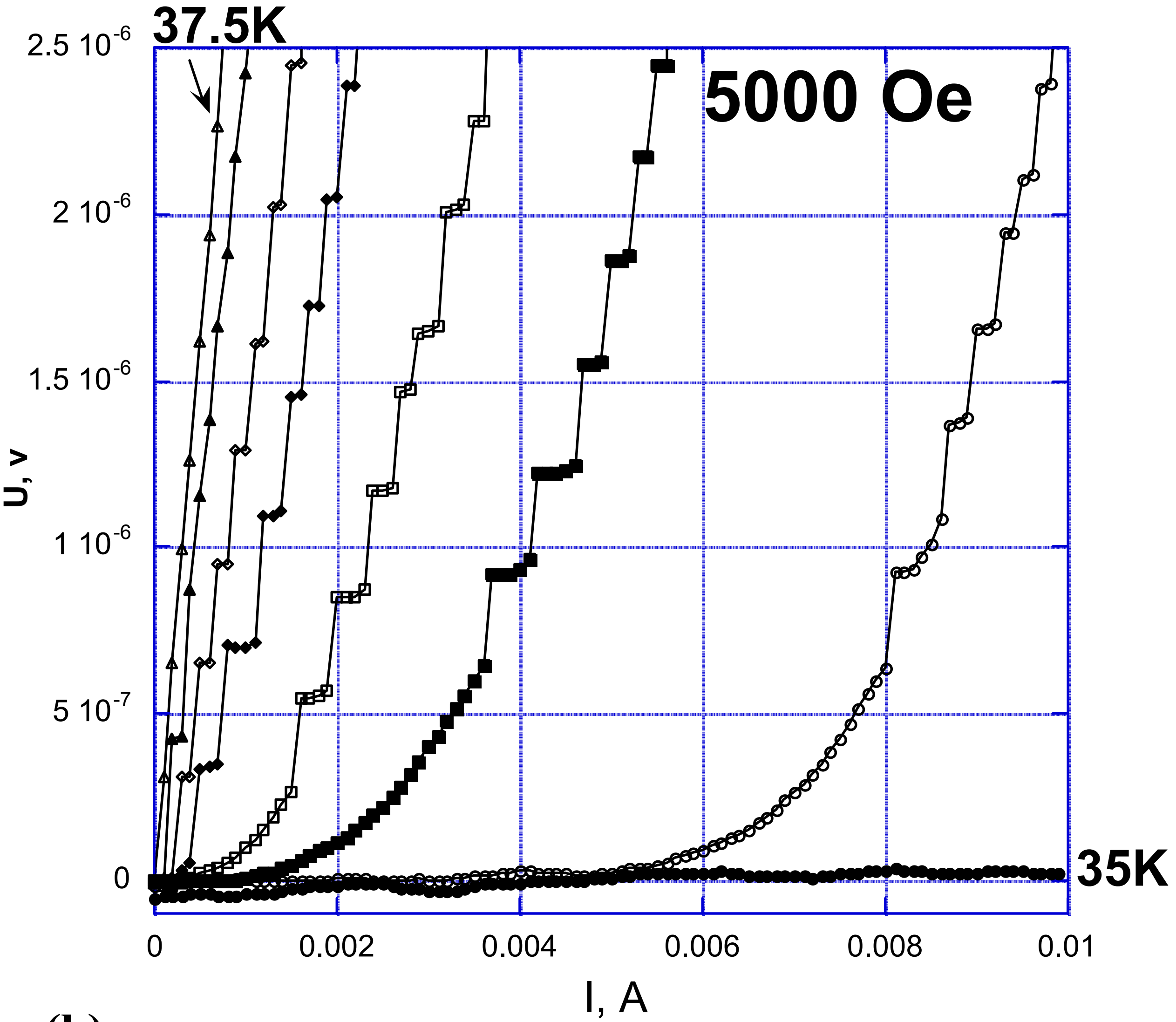

**(b)**

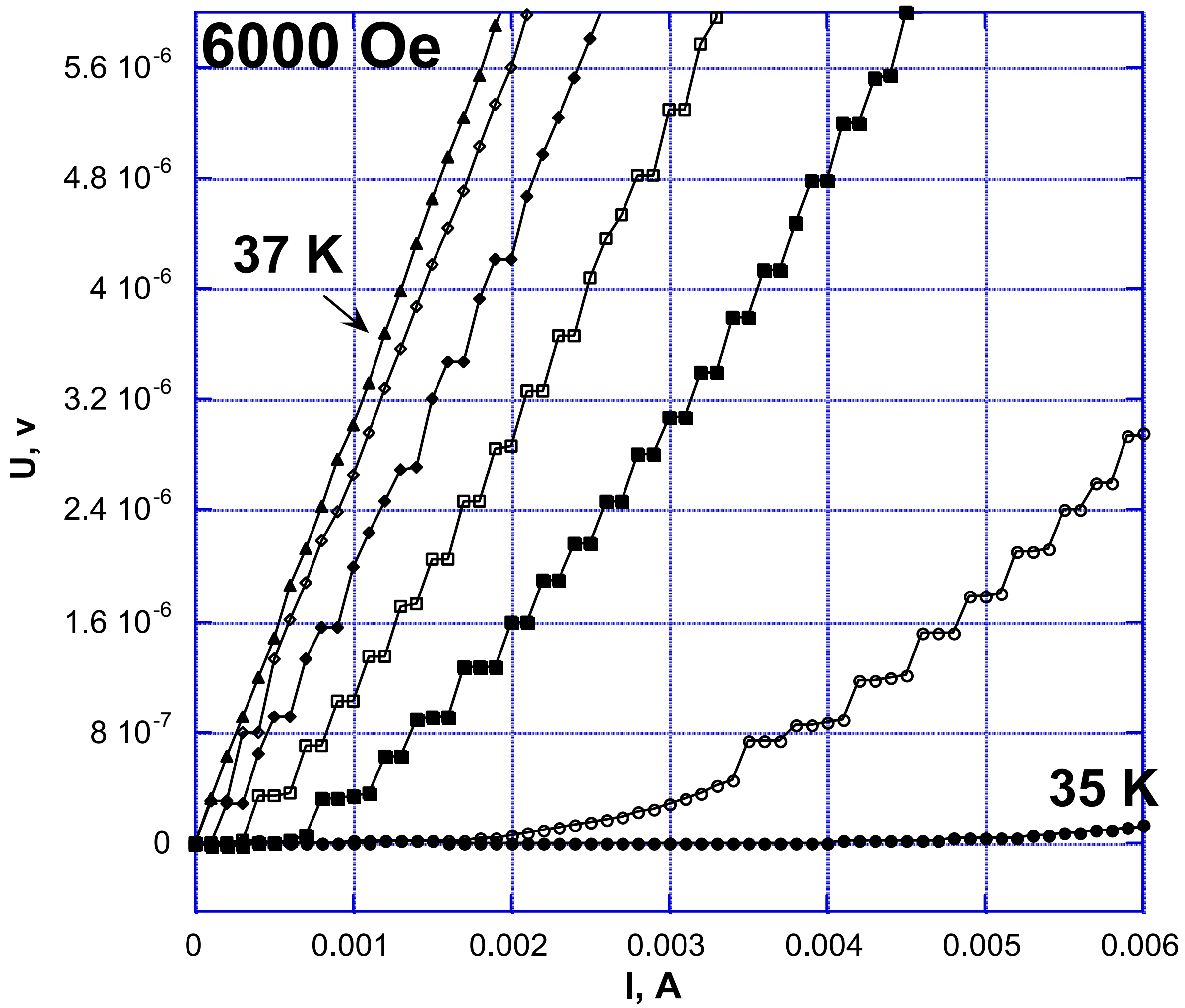

(c)

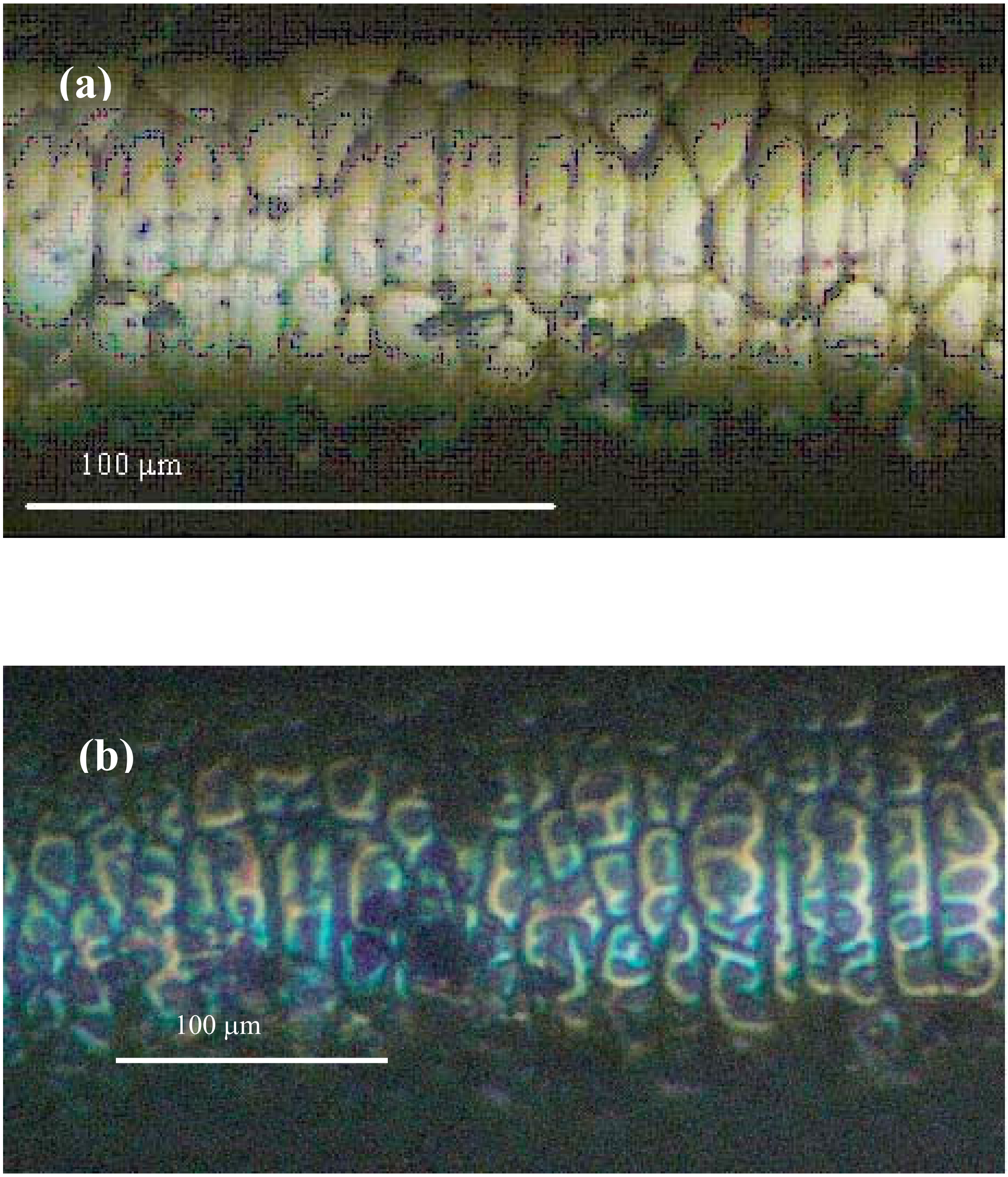

***Fig.** 7:* An optical microscopy image of the boron wire (a) and of a $MgB_2$ wire segment coated with an evaporated ~120 nm gold layer (b).